\documentclass[epjCONF, onecolumn] {svjour}

\usepackage{graphics}

\usepackage[varg]{txfonts} 
\usepackage[latin1]{inputenc}
\session-title{Assembling the Puzzle of the Milky Way}
\begin{document}
\title{The Assembly of the Halo System of the Milky Way as Revealed by
SDSS/SEGUE -- The CEMP Star Connection}
\author{Timothy C. Beers\inst{1},\inst{2}\fnmsep\thanks{\email{beers@noao.edu}} \and Daniela Carollo\inst{3},\inst{4}}
\institute{National Optical Astronomy Observatory, Tucson, AZ, USA \and Michigan State University and JINA, E. Lansing, MI, USA \and Macquarie University, Sydney, AUS \and INAF - Osservatorio Astronomico di Torino, Italy}
\abstract{In recent years, massive new spectroscopic data sets, such as
the over half million stellar spectra obtained during the course of SDSS (in
particular its sub-survey SEGUE), have provided the quantitative detail required
to formulate a coherent story of the assembly and evolution of the Milky Way.
The disk and halo systems of our Galaxy have been shown to be both more complex,
and more interesting, than previously thought.  Here we concentrate on the halo
system of the Milky Way.  New data from SDSS/SEGUE has revealed
that the halo system comprises at least two components, the
inner halo and the outer halo, with demonstrably different characteristics
(metallicity distributions, density distributions, kinematics, etc.). In
addition to suggesting new ways to examine these data, the inner/outer halo
dichotomy has enabled an understanding of at least one long-standing observational
result, the increase of the fraction of carbon-enhanced metal-poor (CEMP) stars
with decreasing metallicity.
}
%
\maketitle
\section{Introduction}\label{intro}

For decades, ``the halo'' of the Galaxy has been thought of as a single entity,
describable in terms of a single metallicity distribution function (MDF), single
density profile, and a simple kinematical distribution that distinguishes it from
the disk system. Indeed, much of the effort since its recognition as an
important component of the Galaxy has concentrated on the effort to better
specify these characteristics \cite{Beers05,Helmi08,Frebel11}. However, over this same
period, numerous clues have emerged that this story is incomplete.  Most of
these clues were obtained in the form of disjoint small-n samples, making it difficult to
``see the forest for the trees.''  This situation has now changed significantly.

Recently, Carollo et al. (2007, \cite{Carollo07}) and Carollo et al. (2010, \cite{Carollo10})
made use of large samples (n $> 30,000$) of ``calibration stars'' from SDSS/SEGUE, in order to directly test the
nature of the halo component of the Galaxy, and elucidate its kinematic and
chemical properties. Their results indicate that the halo consists of two
broadly-overlapping structural components, an ``inner halo" and an ``outer
halo". Note that these labels are not merely descriptors for the regions studied,
but rather are labels for two individual stellar populations. These components
exhibit different MDFs, spatial-density profiles, and stellar orbits, with the 
inner- to outer-halo transition occurring, according to these
authors, at Galactocentric distances of 15-20 kpc. The inner halo was
shown to comprise a population of stars exhibiting a flattened spatial-density
distribution, with an inferred axial ratio on the order of $q_H\sim0.6$, no net
rotation at the level of $\sim$10 km s$^{-1}$, and a MDF peaked at
[Fe/H] $\sim-1.6$. The outer halo comprises stars that exhibit a more spherical
spatial-density distribution, with an axial ratio $q_H\sim0.9$, a clear
retrograde net rotation ($\langle v_{\Phi}\rangle$ $\sim -80$ km s$^{-1}$), a
kinematically ``hotter'' distribution of space velocities than the inner halo,
and a MDF peaked at [Fe/H] $\sim-2.2$. The fundamental demonstration of these
results is contained in the original papers, as well as in \cite{Beers11}, and
will not be repeated here. We note, however, that the dual-halo model for the
Milky Way has received considerable recent theoretical support, based on
simulations for galaxy formation including baryons (as opposed to previous pure
dark-matter-only simulations), e.g., \cite{Font11}, \cite{McCarthy11}, and
\cite{Tissera11}.

In this brief communication, we highlight how recognition of the inner/outer halo
structure of the Milky Way's halo has provided the means for understanding
at least one long-standing ``mystery'' concerning the nature of a specific class of very
metal-poor stars, the so-called carbon-enhanced metal-poor (CEMP) stars.

\section{The CEMP Stars}\label{sec:1}

The CEMP stellar classification was originally defined as the subset of very metal-poor ([Fe/H] $\leq -$2) stars
that exhibit elevated carbon abundances relative to iron, [C/Fe] $>$
+1.0 \cite{Beers05}.  More recently, other criteria have been used, e.g.,
[C/Fe] $> +0.7$ \cite{Aoki07}, which appear to be better supported by the bulk
of the available data.

In the last two decades it has been recognized that roughly 20\% of stars with
[Fe/H] $< -$2.0 exhibit enhanced [C/Fe] ratios (which we refer to as ``carbonicity"), up to several orders of
magnitude larger than the solar ratio \cite{Marsteller05,Lucatello06}. The fraction of CEMP
stars rises to 30\% for [Fe/H] $< -3.0$, 40\% for [Fe/H] $< -3.5$, and 75\% for
[Fe/H] $< -4.0$ (\cite{Beers05}, the fraction would have been 100\%, until
the recent discovery of an SDSS turnoff star with [Fe/H] $\sim-5.0$,
reported by \cite{Caffau11}); definitive explanations for the origin of this increase have
yet to be offered. Regardless of the ultimate reason, these results indicate
that significant amounts of carbon were produced in the early stages of chemical
evolution in the Universe.

There exist a number of sub-classes of CEMP stars, some of which have been
associated with proposed progenitor objects. The CEMP-$s$ stars (those with
$s$-process-element enhancement), for example, are the most commonly observed type to date.
High-resolution spectroscopic studies have revealed that around 80\% of CEMP
stars exhibit $s$-process-element enhancement \cite{Aoki07}. The favored
mechanism invoked to account for these stars is mass transfer of carbon-enhanced
material from the envelope of an asymptotic giant-branch (AGB) star to its
binary companion; it is this surviving companion that is now observed as a
CEMP-$s$ star.

The sub-class of CEMP-$no$ stars (which exhibit no strong neutron-capture-element
enhancements) is particularly prevalent among the lowest metallicity stars (Fe/H
$< -$2.5). Possible progenitors for this class include massive, rapidly
rotating, mega metal-poor ([Fe/H] $< -6.0$) stars, which models suggest have
greatly enhanced abundances of CNO due to distinctive internal burning and
mixing episodes, followed by strong mass loss \cite{Meynet10}. Another
suggested mechanism for the production of the material incorporated into
CEMP-$no$ stars is pollution of the interstellar medium by so-called faint
supernovae associated with the first generations of stars, which experience
extensive mixing and fallback during their explosions \cite{Umeda03}.

The increasing frequency of CEMP stars with declining metallicity, as well as
the suggested increase of the fraction of CEMP stars with increasing distance
from the Galactic plane \cite{Frebel06}, can be explained in the context of an
inner/outer halo dichotomy, and the dominance of {\it different carbon production
mechanisms} (the processes associated with the progenitors of the CEMP-$s$ and
CEMP-$no$ stars) being linked to these two populations.

\section{The Inner/Outer Halo, CEMP-Star Connection}\label{sec:2}

The recent study of Carollo et al. (2011, \cite{Carollo11}) illuminates the intimate connection
between the nature of the halo system and the observed facts concerning CEMP
stars.  Among other results, these authors find that:

\newpage

\begin{figure}[ht]
\hspace{-1.5cm}
\resizebox{1.2\columnwidth}{!}{\includegraphics{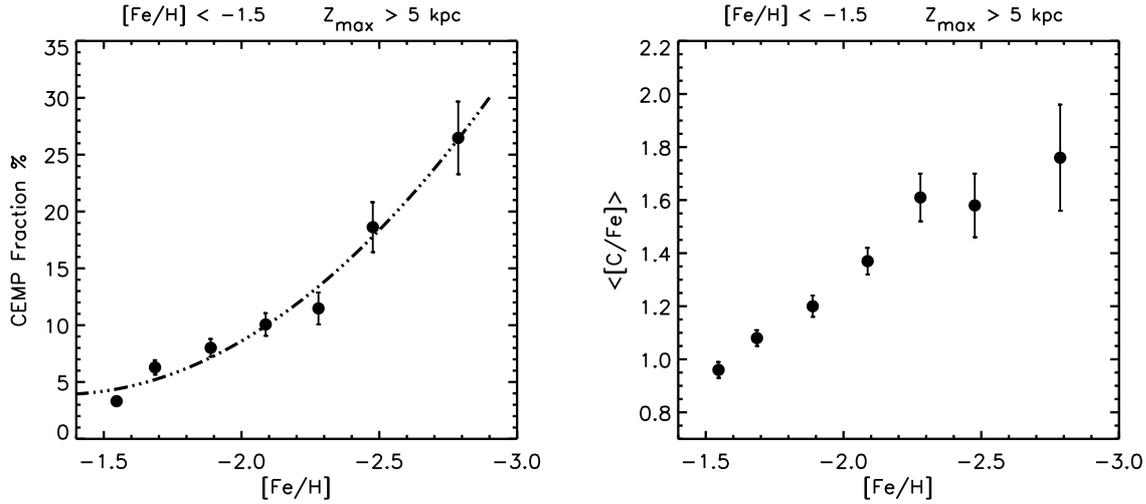}}
\caption{Left panel: Global trend of the CEMP fraction, as a function of [Fe/H],
for the low-metallicity SDSS calibration stars with Z$_{max}$  $> 5$ kpc. Each bin
of metallicity has width $\Delta$[Fe/H] = 0.2 dex, with the exception of the
lowest-metallicity bin, which includes all stars with [Fe/H] $< -2.6$. The error
bars are evaluated with the jackknife method.  A second-order polynomial fit
to the observed distribution is shown by the dot-dashed line.
Right panel: Global trend of the mean carbonicity, $\langle$[C/Fe]$\rangle$, as
a function of [Fe/H]. Only those stars with detected CH G-bands are used. The
bins are the same as used in the left panel. Errors are the standard error in
the mean.}
\label{fig:1} 
\end{figure}

\begin{itemize}

\item \emph{Almost all} of the CEMP stars are located in the halo components
of the Milky Way.

\medskip

\item At low metallicities ([Fe/H] $< -1.5$), the distribution of derived
distances and rotational velocties for the CEMP and non-CEMP stars differ
significantly from one another.

\medskip

\item In the low-metallicity regime, [Fe/H] $< -$1.5, and at vertical
distances Z$_{max} >$ 5 kpc (where Z$_{max}$ refers to the maximum distance from
the Galactic plane reached during the course of a stellar orbit),
a continued increase of the fraction of CEMP stars
with declining metallicity is found (Figure 1, left panel). A second-order
polynomial provides a good fit to the CEMP star fraction as a function of [Fe/H].

\medskip

\item In the same metallicity regime, the mean carbonicity, 
$\langle$[C/Fe]$\rangle$, increases with declining metallicity (Figure 1, right
panel).

\medskip

\item In the low-metallicity regime, for both $-2.0 <$ [Fe/H] $< -$1.5
and [Fe/H] $< -$2.0, Carollo et al. find a clear increase in the CEMP star
fraction with distance from the Galactic plane, $|$Z$|$ (Figure 2). At these low
metallicities, significant contamination from the disk populations is unlikely,
and would only reasonably have a small effect for the bin with 0 kpc $<$ $|$Z$|$
$<$ 4 kpc and $-2.0 <$ [Fe/H] $< -$1.5; this is an observed property of the halo
system. \textit{Such a result would be difficult to understand in the context of a
single halo population}.

\medskip

\item At low metallicity, $-2.0 <$ [Fe/H] $< -1.5$, where the dominant component
is the inner halo, a kinematic decomposition of this population from the outer-halo population reveals
no significant difference in the CEMP star fraction
between the inner/outer halo. In contrast, at very low metallicity, $-2.5 <$
[Fe/H] $< -2.0$, where the dominant component is the outer halo, there is
evidence for a significant difference in the CEMP star fraction between the
inner halo and the outer halo (Figure 3); the fraction of CEMP stars associated with 
the outer halo in this metallicity interval is roughly twice that associated
with the inner halo. This difference is also confirmed
in the lowest metallicity bin, $-3.0 <$ [Fe/H] $< -2.5$, even though it is less
remarkable, due to the smaller numbers of stars and proportionately larger error bars.
Carollo et al. conclude that the difference in CEMP frequency at
very low metallicity is driven \textit{not by metallicity itself, but rather, by the
stellar populations present.}

\end{itemize}

\begin{figure}[h]
\resizebox{0.9\columnwidth}{!}{\includegraphics{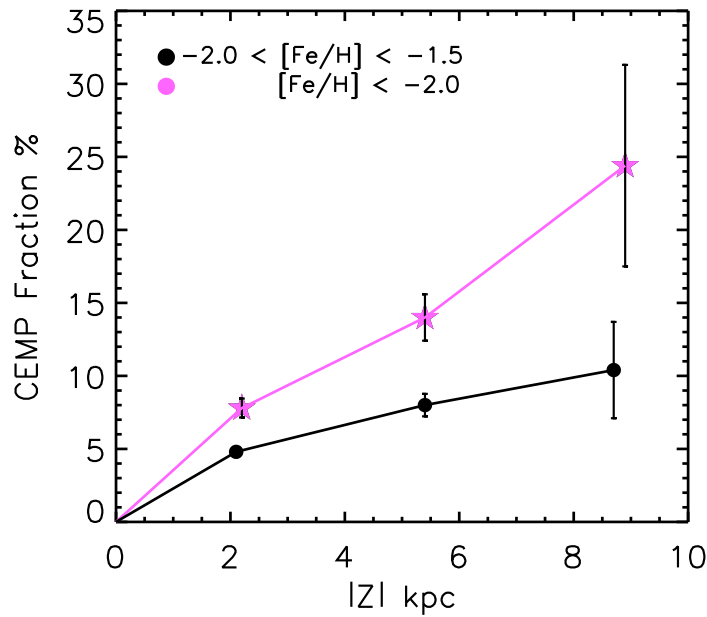}} 
\caption{Global trend of the
CEMP star fraction, as a function of distance from the Galactic plane, $|$Z$|$,
for the low-metallicity stars of the Extended Sample described by \cite{Carollo11}.
The stars with $-$2.0 $<$[Fe/H] $<-1.5$ are shown as filled black circles; those with [Fe/H] $< -2.0$are shown as magenta stars.  Each bin has a width of $\Delta$$|$Z$|$ = 4 kpc, with the
exception of the last bin, which cuts off at 10 kpc, the limiting distance
of the Extended Sample.  Note the clear dependence of
CEMP fraction on height above the Galactic plane, for both metallicity regimes.}

\label{fig:2}
\end{figure}

\begin{figure}[h]
\hspace{0.5cm}
\resizebox{0.9\columnwidth}{!}{\includegraphics{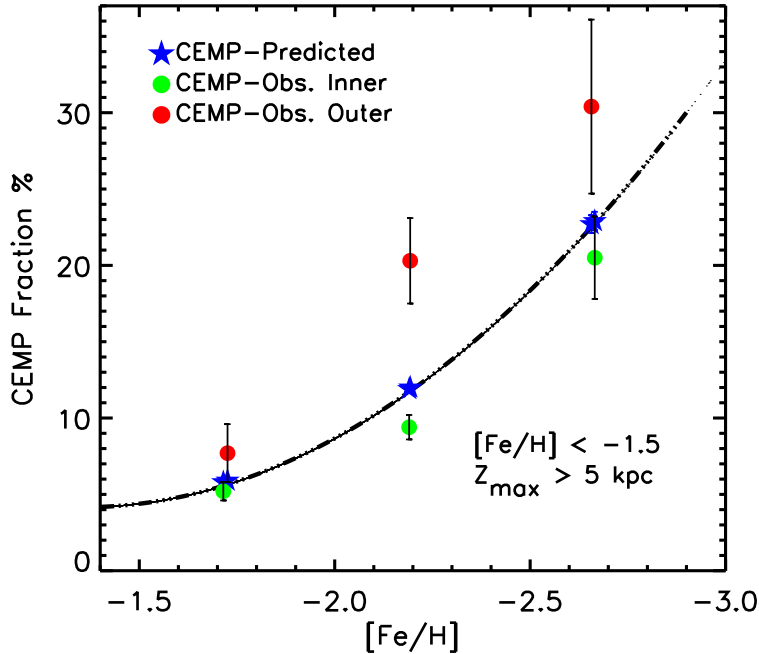}} 
\caption{Global trend of the CEMP star fraction as a function
of [Fe/H] for the low-metallicity stars of the Extended Sample with
Z$_{max}$ $> 5$ kpc.  The dashed curve is a second-order polynomial fit representing
the global trend. The blue filled stars represent the predicted values of the
CEMP fractions in each bin of metallicity, having a width of $\Delta$[Fe/H]
= 0.5 dex. The green and red filled circles show the location of the observed
CEMP star fractions for the inner halo and outer halo, respectively, based on
the kinematic deconvolution method described by \cite{Carollo11}.}

\label{fig:3}
\end{figure}

\section{Implications for Galaxy Formation}\label{sec:3}

The large fractions of CEMP stars in both halo components indicates that
significant amounts of carbon were produced in the early stages of chemical
evolution in the Universe. The observed contrast of CEMP star fractions in the
inner- and outer-halo populations strengthen the picture that the halo components had
different origins, and supports a scenario in which the outer-halo component has
been assembled by the accretion of small subsystems. In this regard, it is
interesting that the MDF of the inner halo (peak at [Fe/H] $\sim -$ 1.6) is in
the metallicity regime associated with the CEMP-$s$ stars, which are primarily
found with [Fe/H] $> -2.5$, while the MDF of the outer halo (peak at [Fe/H]
$\sim$ $-$2.2), might be associated with the metallicity regime of the CEMP-$no$
stars, which are primarily found with [Fe/H] $< -2.5$.

The fact that, in the metal-poor regime, the outer halo exhibits a fraction of
CEMP stars that is larger than the inner halo suggests that multiple sources of
carbon, besides the nucleosynthesis of AGB stars in binary systems, were present
in the pristine environment of the outer-halo progenitors (lower mass sub-halos).
These sources could be the fast massive rotators and/or the faint supernovae
mentioned above. If the CEMP stars in the outer halo are predominantly
CEMP-$no$ stars (which has yet to be established), it might suggest that
non-AGB-related carbon production took place in the primordial mini-halos. The
predominance of CEMP-$s$ stars in the inner halo, if found, would suggest that the
dominant source of carbon was the nucleosynthesis in AGB stars in a binary
system. This would place important constraints on the primordial IMF of the
sub-systems responsible for the formation of the two halo components.

Recent efforts to model, from the population synthesis standpoint, the fractions
of observed CEMP stars in the halo have struggled to reproduce results as high
as 15-20\% for metallicities [Fe/H] $< -2.0$ \cite{Izzard09,Pols10}. Note, however,
that these predictions are based solely on carbon production by AGB stars.
While such calculations may prove meaningful for the inner-halo population, they
may not be telling the full story for carbon production associated with the
progenitors of the outer-halo population. The lower CEMP fraction in the inner
halo may relieve some of the tension with current model predictions.

The fact that the CEMP star fraction exhibits a clear increase with $|$Z$|$
suggests that the relative numbers of CEMP stars in a stellar population is
not driven by metallicity alone. The proposed coupling of the cosmic microwave
background to the IMF \cite{Larson05,Tumlinson07} is one mechanism for imposing
a temporal dependence on the IMF. This effect, coupled with chemical evolution
models, predicts that the CEMP fraction would be expected to increase as the
metallicity decreases, but with similar metallicity regimes forming carbon
according to the expected yields of the predominant mass range available at that
time. In the hierarchical context of galaxy formation, star-forming regions are
spatially segregated, and their chemical evolution can proceed at different
rates, with stars at the same metallicity forming at different times. Depending
on the source(s) of carbon, this trend could lead to a spatial variation of the
CEMP fraction at the same metallicity, increasing in older populations and
decreasing in younger ones.

\section{The Path Forward}\label{sec:4}

We have used this article to illustrate one application of the dual-halo model
to understand the formerly unexplained increase in the fraction of CEMP stars
with declining metallicity, demonstrating that it arises from the differing 
nature of the stellar populations associated with the inner/outer halo.

Further exploration of the CEMP stars, in particular the CEMP-$no$ sub-class, which
may well have arisen from non-AGB sources of carbon production, should
prove of great interest.  Connections of such stars with chemical production in early epochs are now being made.
For example, the recently reported high-redshift ($z = 2.3$), extremely metal-poor Damped Lyman-$\alpha$
system (\cite{Cooke11}, [Fe/H] $\sim -3.0$) exhibits enhanced
carbonicity ([C/Fe] $= +1.5$) and other elemental abundance signatures that are quite 
similar to those of the CEMP-$no$ stars, and which \cite{Kobayashi11} associate with production by faint supernovae.
It is also of interest that evidence for strong carbon production in the early Universe
has been reported \cite{Matsuoka11}, based on analysis of the optical spectrum of the most 
distant known radio galaxy, TN J0924-2201, with $z = 5.19$.  When viewed in this light,
the CEMP-$no$ stars may well prove to be our local touchstone to the abundance patterns produced
by the very FIRST generations of stars, an exciting prospect!

\end{document}